\documentclass[10pt,twocolumn]{article}
\usepackage[utf8]{inputenc}
\usepackage[T1]{fontenc}
\usepackage{amsmath,amsfonts,amssymb}
\usepackage{graphicx}
\usepackage{booktabs}
\usepackage{array}
\usepackage{multirow}
\usepackage{url}
\usepackage{geometry}
\usepackage{float}
\usepackage{caption}
\usepackage{subcaption}
\usepackage{algorithm}
\usepackage{algorithmic}
\usepackage{hyperref}
\usepackage{multicol}
\usepackage{balance}
\usepackage{dblfloatfix}

\usepackage{titlesec}
\titleformat{\section}
  {\normalfont\large\bfseries}{\thesection}{1em}{}
\titleformat{\subsection}
  {\normalfont\normalsize\bfseries}{\thesubsection}{1em}{}
\titleformat{\subsubsection}
  {\normalfont\small\bfseries}{\thesubsubsection}{1em}{}
\titlespacing\section{0pt}{8pt plus 2pt minus 2pt}{4pt plus 2pt minus 2pt}
\titlespacing\subsection{0pt}{6pt plus 2pt minus 2pt}{3pt plus 2pt minus 2pt}
\titlespacing\subsubsection{0pt}{4pt plus 2pt minus 2pt}{2pt plus 2pt minus 2pt}

\usepackage{enumitem}
\setlist{nosep, leftmargin=*, before=\vspace{-0.5\baselineskip}, after=\vspace{-0.5\baselineskip}}

\title{DUALRec: A Hybrid Sequential and Language Model Framework for Context-Aware Movie Recommendation}

\author{Yitong Li, Raoul Grasman\\
University of Amsterdam\\
\texttt{yitong.li3@student.uva.nl}}

\date{}

\begin{document}

\twocolumn[
\begin{@twocolumnfalse}
\maketitle

\begin{abstract}
The modern recommender systems are facing an increasing challenge of modelling and predicting the dynamic and context-rich user preferences. Traditional collaborative filtering and content-based methods often struggle to capture the temporal patternings and evolving user intentions. While Large Language Models (LLMs) have gained gradual attention in recent years, by their strong semantic understanding and reasoning abilities, they are not inherently designed to model chronologically evolving user preference and intentions. On the other hand, for sequential models like LSTM (Long-Short-Term-Memory) which is good at capturing the temporal dynamics of user behaviour and evolving user preference over time, but still lacks a rich semantic understanding for comprehensive recommendation generation.

In this study, we propose DUALRec (Dynamic User-Aware Language-based Recommender), a novel recommender that leverages the complementary strength of both models, which combines the temporal modelling abilities of LSTM networks with semantic reasoning power of the fine-tuned Large Language Models. The LSTM component will capture users' evolving preference through their viewing history, while the fine-tuned LLM variants will leverage these temporal user insights to generate next movies that users might enjoy. Experimental results on MovieLens-1M dataset shows that the DUALRec model outperforms a wide range of baseline models, with comprehensive evaluation matrices of Hit Rate (HR@k), Normalized Discounted Cumulative Gain (NDCG@k), and genre similarity metrics. This research proposes a novel architecture that bridges the gap between temporal sequence modeling and semantic reasoning, and offers a promising direction for developing more intelligent and context-aware recommenders.
\end{abstract}

\vspace{0.5cm}
\end{@twocolumnfalse}
]

\section{Introduction}

Recommendation systems, according to the definition by IBM (2024), is an artificial intelligent (AI) system that suggests items to users. And it plays a crucial role in improving user engagement by predicting user preferences based on historical interactions. Traditional approaches like collaborative filtering (CB, Sarwar et al., 2001) and content-based filtering (CBF, Musto et al., 2022) have provided valuable foundations for modern recommenders. CF uses historical user-item interaction data to predict the user preferences, while CBF settled on item features to match relevant content with users (Wang et al., 2024). However, despite their widespread research and adoption, these approaches do fall short in two tasks: (1) capturing the temporal dynamics of user behavior as preferences evolve overtime, and (2) understanding the semantic richness behind user-item interactions, such as underlying themes, genres, or user intent shifts (Xu et al., 2025).

In recent years, there has been increasing research attention in improving the recommendation systems with more complex models that are capable of understanding user intent and context (Xu et al., 2025). Traditional models often assumes user preferences are based on past behaviour and static, while recent research has started to leverage contextual information including temporal patterns like user sentiment (Yu et al., 2021), event information and emotional state (Liu et al., 2024; Liu et al., 2023) to structure dynamic recommenders.

Large Language Models (LLMs) in this case have shown promising performance in understanding the complex user context and inputs (Yu et al., 2024). With LLMs capability in encoding vast amounts of common sense knowledge and understanding of complex user behaviours in complex scenarios, especially in multifaceted scenarios involving contextual, situational or emotional factors (Wang et al., 2018). Therefore, it is worthwhile to explore how LLMs can benefit recommender systems to better represent users and items and enhance the performance of recommender accuracy (Wang et al., 2018).

However, with recent studies exploring the integration of LLMs for more personalized recommenders with natural language understanding (Ren et al., 2024), and context-aware recommenders to incorporate real-time factors like social context and sentiments, there is still lack of robust and reliable model for adapting and predicting the user intent changes in a timely manner (Bakhshizadeh, 2024). Overall, there is a need for a recommender that can dynamically interpret user preference from sparse data and adapt to recommendations in real-time.

To address these challenges, the current research proposes a novel hybrid approach, named DUALRec (Dynamic User-Aware Language-based Recommender). In our approach, the LSTM model is first employed to capture the dynamic intent of user preference based on their recent activities. Then the output from LSTM and user movie watching history will then structured together as a natural language prompt template to feed into our choice of four LLMs from DeepSeek and Mistral models (DeepSeek-AI, 2024a; Mistral AI, 2023a, 2023b) as variants to predict the next movie the user might be interested in. And we will evaluate the performance of for DUALRec variants with the usage of four different LLMs using the Hit Rate at n (HR@(1,5)), Normalized Discounted Cumulative Gain at n (NDCG@(1,5)) and Jaccard Genre Similarity matrices.

\section{Related Work}

\subsection{Classic Recommenders}

Recommender systems have evolved significantly over the past decades, from the classic collaborative filtering (CF) that recommends items based on similar users' preference (Sarwar et al., 2001). Content based filtering (CB) (Musto et al., 2022) that recommends items with similar attributes to those previously liked by users. And to more complicated neural network techniques such as neural collaborative filtering (NCF) (He et al., 2017) which leverages neural networks to capture complex, non-linear user–item interactions.

And the Wide \& Deep method (Cheng et al., 2016) which jointly trains linear models (wide) and deep neural networks for better memorization of feature interactions and generalization to unseen feature combinations. However, those classic recommenders are more static recommendation models (Gao et al., 2021), where they treat user-item interactions as static and unchanging overtime, so they may fail to capture the dynamic intention and preference changes in real time user item interactions (Covington et al., 2016; Jagerman et al., 2019; Gao et al., 2021).

\subsection{Intent Modelling via Sequential and Recurrent Methods}

Sequential recommendations, on the other hand, have been proposed as a strategy for modelling the evolving user preference, especially when the order and timing of user interactions are critical. Long Short-Term Memory (LSTM) networks in this case, as a type of recurrent neural network(RNN), have been widely used for this tasks with their ability to capture long-term dependencies and user behavioural patterns over time (Hidasi et al., 2015; Donkers et al., 2017). Unlike classical CF methods that treat user interactions as an unordered set, without modelling the order to timing of actions, LSTM networks can extract dynamic user profiles by learning from sequential patterns in behavior. Rather than relying on static preferences, LSTMs adapt to changes in interest over time by processing ordered user sessions and updating internal representations accordingly (Hidasi et al., 2015; Quadrana et al., 2017).

In the context of our current movie recommendation tasks, LSTMs are particularly effective when the user interaction data is accompanied by timestamps (Harper \& Konstan, 2015). These temporal signals allow the model to learn not only which items users consume but also when they consume them to reveal time sensitive interests or evolving movie genre preferences. However, despite their strength, LSTM do face several limitations. With strong modelling of sequence dynamics, they are still restricted to ID-based predictions and lack the ability to reason about semantic relationships between user interacted items (Sun et al., 2019). These limitations thus highlighted the need to combine sequential learning with models capable of semantic reasoning and knowledge generalization.

\subsection{Large Language Models in Recommendation Systems}

To address the limitation of traditional and sequential recommendation model, recent studies have introduced Large Language Models (LLMs) with their ability to process complex user input and content in the form of queries, comments and reviews (Yu et al., 2024; Wang et al., 2024), then extracting a deeper understanding of the needs of users. So, the current approaches will leverage LLMs in recommendation tasks by using natural language prompts, which is a manually constructed natural language template that encodes the recommendation problem. For example, a prompt might read: ``The user has watched 'Inception', 'The Matrix', and 'Interstellar'. What other movies would they enjoy?'' The LLM is then asked to generate one or more recommendations in plain text.

This prompt-based paradigm has been shown to be effective in extracting relevant and personalized suggestions, even without additional training. A representative example of this approach is the Zero-Shot Next-Item Recommendation (NIR) framework introduced by Wang and Lim (2023). In their study, the authors use frozen LLMs (e.g., GPT-3) to perform recommendations through a structured three-step prompting protocol. First, the system identifies the user's overall preferences; second, it selects a representative subset of previously watched items; and third, it constructs a natural language prompt that is used to query the LLM for a ranked list of ten recommendations, generated in structured text format.

The generated outputs are parsed from text and evaluated against ground-truth data using Hit Rate (HR@10). Where, the HR@n (Hit Rate at n) measures the percentage of test cases where the correct next movie appears among the top-n recommendations. On the MovieLens 100K dataset, this zero-shot prompting strategy achieved lower but similar performance compared to several well-established baselines, including FPMC (Rendle et al., 2010), GRU4Rec (Hidasi et al., 2015), and SASRec (Kang \& McAuley, 2018). The study demonstrates that LLMs, when guided by carefully designed natural language prompts and contextual cues, can function as effective recommenders without the need for gradient-based fine-tuning.

This study design directly influenced multiple aspects of our own model structure and evaluation protocol. While Wang and Lim (2023) frame the recommendation as a pure zero-shot generation task using fixed LLMs, we would extend this idea by incorporating the parameter-efficient fine-tuning (PEFT, Mangrulkar et al., 2022), which allows LLMs to be adapted to specific tasks (i.e. movie recommendation) using a small number of trainable parameters while keeping the majority of the model frozen. Specifically, we would adopt the Low-Rank Adaptation (LoRA) technique in PEFT, which introduces low-rank modifications of the weight matrices in the model's attention layers, enabling efficient training and deployment (Hu et al., 2021; Hugging Face, 2023).

\section{Dataset, Coding and LLM Selection}

The study will utilize the well-established MovieLens 1M dataset (Harper \& Konstan, 2015) which is a stable benchmark dataset with 1 million ratings from 6000 users on 4000 movies. Each rating entry includes user ID, movie ID, movie titles, rating value (1-5), genres and timestamp. To ensure a sufficient signal for learning temporal patterns, the dataset was filtered to include only the top 1,000 most frequently watched movies, allowing the model to focus on movies with adequate user interaction data for training. Each movie is represented through three complementary modalities: (1) unique movie IDs embedded into 128-dimensional vectors, (2) tokenized movie titles using Keras Tokenizer, and (3) multi-label genre information. The dataset was then split into training (70\%), validation (15\%), and test (15\%) sets at the user level to ensure no user history leakage between splits, and simulate real-world scenarios where models are trained on past data and evaluated on future data.

Our coding and training were carried out in Google Colab Pro using an A100 GPU with 40GB GPU RAM in Python 3 language. In terms of LLM selection, due to the GPU resource limit of 40GB, it was not feasible to fine-tune DeepSeek V3 and Mistral 8B locally. Therefore, we choose the DeepSeek-R1-Distill-Qwen-1.5B model, and the Mistral 7B model as a high-capacity alternative to the DeepSeek V3 and Mistral 8B model, as they share architectural similarities including dense causal attention, sliding window mechanisms and strong ability to follow instructions and generate contextually grounded output (DeepSeek-AI, 2024b, 2024c; Mistral AI, 2024a, 2024b).

\section{Methodology}

This part introduces the comprehensive methodology for DUALRec, a novel hybrid recommendation system that addresses the fundamental limitations of existing approaches by strategically combining temporal sequence modelling with semantic reasoning.

\subsection{Framing the Current Recommendation Task}

In a movie recommendation scenario, each user interacts with a series of movies over time. These interactions form a sequential viewing history, where each entry records the specific movie a viewer watched along with the corresponding timestamp. This sequence reflects both the user's evolving interests and the temporal patterns in their viewing behaviour.

The objective of the recommendation task is to predict the next movie a user is most likely to watch, given their prior viewing history. Specifically, as shown in Figure 1, for each user, we aim to model the relationship between their past sequence of movie watching history and predict the likely next movie they will watch through our recommender model. This task can be seen as a conditional sequence prediction problem, where the model learns to associate patterns in historical sequence with likely future outcomes.

\begin{figure}[htbp]
\centering
\includegraphics[width=\linewidth]{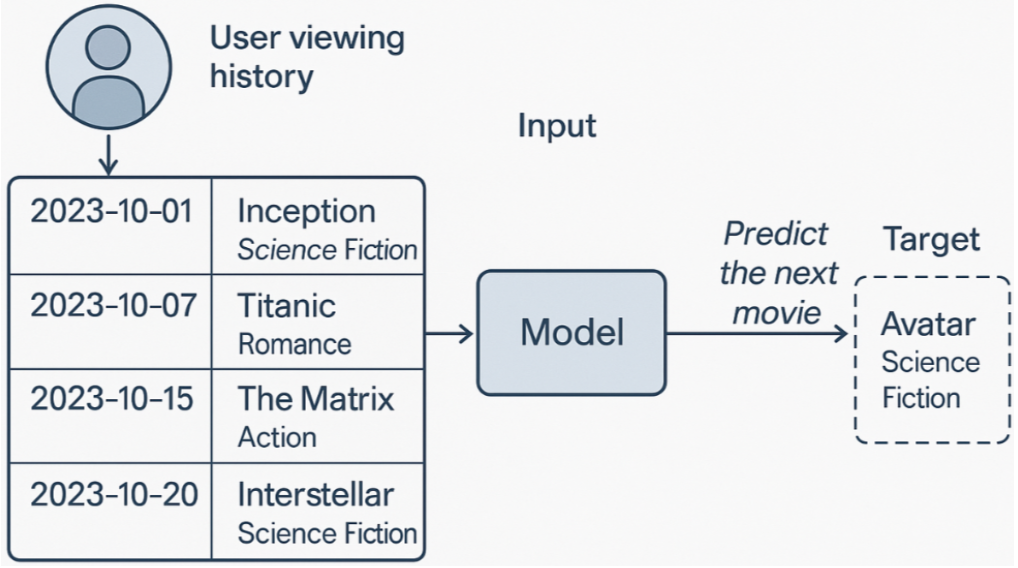}
\caption{The current recommendation task in a workflow.}
\label{fig:recommendation_task}
\end{figure}

\subsection{DUALRec Theoretical Framework}

The Dynamic User-Aware Language-based Recommender (DUALRec) framework predicts recommendations from a three stage process. In the first stage, we trained an LSTM model on users' movie viewing sequences to perform a sequence prediction task. The LSTM predicts the next movie a user is most likely to watch. This predicted movie, along with the user's recent viewing history, is then used to construct a natural language prompt. This prompt is then fed into the DeepSeek V3, and Mistral 8B models in the second stage, which generates a response with a list of semantically relevant movie titles with release years and genre information as personalized recommendations.

And in the third stage, we further enhanced the model by adopting the LoRA-based PEFT strategy to fine-tune two other open-source language models: DeepSeek-R1-Distill-Qwen-1.5B and Mistral-7B. We evaluate the recommendation quality using several metrics: Hit Rate at k (HR@1, HR@5), Normalized Discounted Cumulative Gain (NDCG@1, NDCG@5), and genre-level Jaccard similarity, which measures the overlap in genre tags between predicted and actual next movies.

\subsubsection{Stage 1: Sequential Behaviour Modelling}

The first stage of the DUALRec framework involves learning user behaviour patterns over time using a multimodal two-layer Long Short-Term Memory (LSTM) network. The user-item interaction data was first grouped by user and sorted chronologically by timestamp to construct a sequential dataset. A sliding window approach was applied to construct sequences of 30 consecutively watched movies for each user, with the 31st movie serving as the ground truth target that would be used to evaluate recommendation accuracy. Each movie within the sequence is represented by its unique movie ID, along with its tokenized title and genre features, creating a multimodal input at each sequence timestep.

Tokenized Movie Titles are processed using Keras tokenizer with a vocabulary capped at 5,000 words. Each title is converted into a sequence of up to 10 tokens. The 10-token limit for titles balances computational efficiency with semantic richness, accommodating the typical length distribution of movie titles while enabling batch processing. Each token is embedded using a trainable 64-dimensional word embedding layer, and the token sequence is passed through a global average pooling layer to generate a fixed 64-dimensional representation for each title. This process is repeated for all 30 time steps in the user's viewing history.

Genre Features are encoded as binary multi-hot vectors across 18 possible genre categories ('Action' 'Adventure' 'Animation' "Children's" 'Comedy' 'Crime' 'Documentary' 'Drama' 'Fantasy' 'Film-Noir' 'Horror' 'Musical' 'Mystery' 'Romance' 'Sci-Fi' 'Thriller' 'War' 'Western'). A movie may belong to multiple genres, and each genre vector is passed through as a 64-dimensional fully connected dense layer with ReLU activation to reduce its dimensionality and introduce nonlinearity.

For each time step, the three feature vectors (movie embedding, title representation, and genre) are concatenated into a unified feature vector. These concatenated vectors form a sequence of scale $30 \times d$, where $d$ is the total combined feature dimensionality. This multimodal sequence input is then passed through a two-layer LSTM:

\begin{figure}[htbp]
\centering
\includegraphics[width=\linewidth]{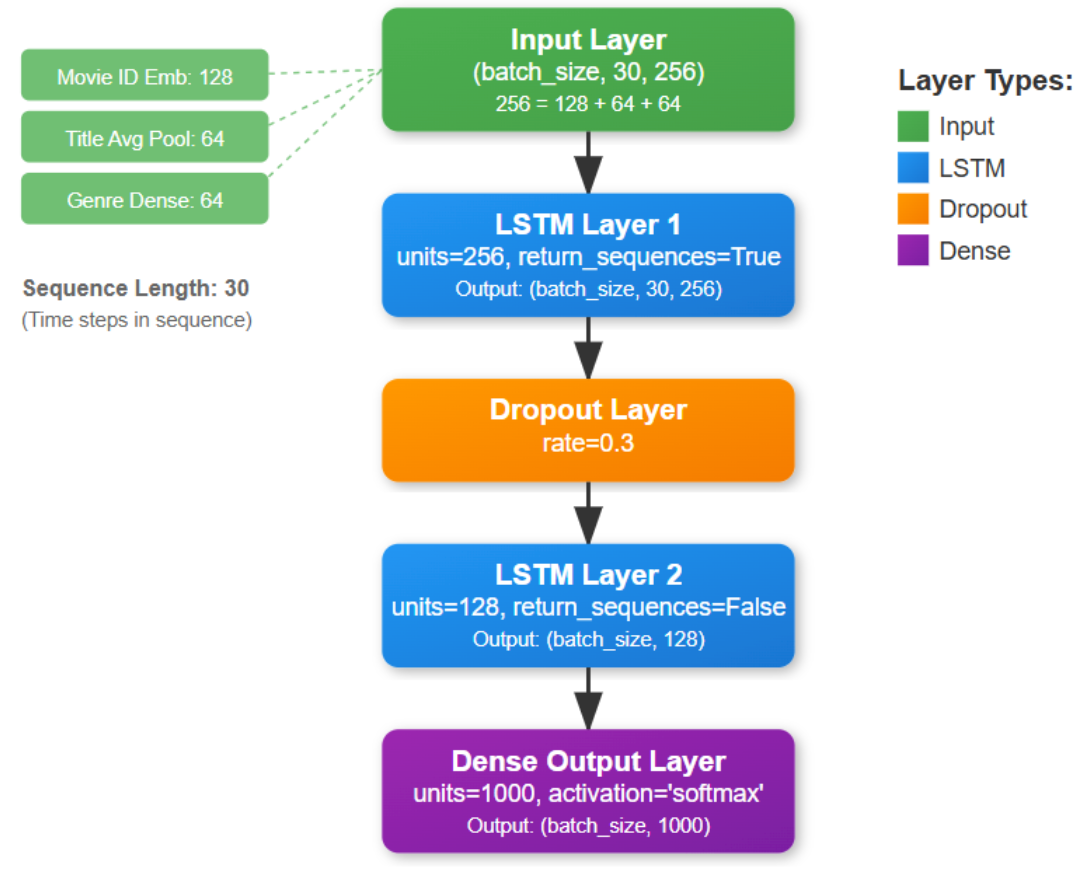}
\caption{Two-layer LSTM architecture with input shape (30, 256), combining embeddings from movie ID, title, and genre. It outputs top-k movie predictions via a 128-unit LSTM, dropout (0.3), and dense SoftMax over 1,000 classes.}
\label{fig:lstm_architecture}
\end{figure}

The first LSTM layer has 256 hidden units and returns sequences, enabling it to learn local, short-range dependencies between adjacent movie choices. The second LSTM layer has 128 units and returns the final hidden state, encoding a global representation of the user's long-term preferences. A dropout rate of 0.3 is applied to both layers to prevent overfitting. The final LSTM output is passed through a dense layer and then a softmax layer that produces a probability distribution over the 1,000 most frequent movies. The top-1 prediction from this distribution is representing the most likely next movie based on temporal dynamic prediction alone.

To prepare the LSTM output into Stage 2, we have the predicted movie ID mapped back to its corresponding movie title. And this title is then incorporated into a natural language prompt that is later fed into the large language model (LLM), enabling semantic generation conditioned on both viewing history and the LSTM's predicted next item.

\subsubsection{Stage 2: Language-Based Contextual Reasoning}

In the second stage, DUALRec enhances its recommendation performance using a Large Language Model (LLM) to perform free-form natural language generation tasks. This process builds on the output provided by the Stage 1 LSTM model, combining temporal reasoning with rich, semantic recommendation via prompt-based generation.

Each prompt is dynamically generated by combining three components including the users' most recently watched five movies ordered by timestamp, the top-1 predicted movie from LSTM model (Stage 1), and the standardized instruction template written in natural language to guide LLM to generate coherent recommendations.

\textit{\textbf{Example Prompt to LLM:}}

\textit{Below is a user's movie watching history:}

\textit{- Bug's Life, A (1998) (Animation, Children's, Comedy)}

\textit{- Antz (1998) (Animation, Children's)}

\textit{- Hercules (1997) (Adventure, Animation, Children's, Comedy, Musical)}

\textit{- Mulan (1998) (Animation, Children's)}

\textit{- Pocahontas (1995) (Animation, Children's, Musical, Romance)}

\textit{Based on this, the system (LSTM) recommends: Mission to Mars (2000).}

\textit{Now, as a helpful assistant, recommend 3 more full movie titles with release years and genres that this user would likely enjoy next.}

The prompt is then passed through LLM using an OpenRouter API call. The model will return a generated ranked list of 3 additional movie titles. These recommendations are expected to be semantically cohesive, informed by movie themes, user viewing context and genres.

\textit{\textbf{Example LLM Recommendations:}}

\textit{Based on the user's preference for animated, family-friendly films with adventurous and musical elements, here are three recommendations that align with their viewing history:}

\textit{- Tarzan (1999) Genres: Animation, Adventure, Children's, Musical}

\textit{- The Emperor's New Groove (2000) Genres: Animation, Adventure, Children's, Comedy}

\textit{- Lilo \& Stitch (2002) Genres: Animation, Children's, Comedy, Science Fiction}

\textit{Based on this, the system (LSTM) recommends: Mission to Mars (2000).}

\textit{Now, as a helpful assistant, recommend 3 more full movie titles with release years and genres that this user would likely enjoy next.}

\subsubsection{Stage 3: Language Model Fine-Tuning and Post-Generation Optimization}

In the third stage, we focus on enhancing the semantic quality and contextual alignment of recommendations through language model fine-tuning and post-generation optimization. We expect to refine recommendation outputs by conditioning on both user history and LSTM predictions, followed by an semantic re-ranking step to further align generated outputs with user intent.

We still leverage the sequential nature of each user's movie viewing history from the MovieLens 1M dataset. For every user in the fine-tuning training dataset, their interactions were first sorted chronologically, then we split each user's data into two parts: the earlier portion of viewing history (excluding the last five movies) was used as contextual input, then the last five movies as ground truth window. From this window, three titles are randomly selected as the target output, providing partial supervision while leaving the remaining movies for later evaluation process.

The goal of fine-tuning LLM is to improve the semantic quality and personalization of recommendations. Specifically, we aim to teach the LLM to generate movie suggestions that align with both a user's recent preference and the behavioural signal from LSTM, not just hallucinate generic suggestions. Each example in the fine-tuning dataset follows an instruction–input–output schema that frames the task as instruction-following text generation. The input prompt is constructed from the user's five most recent movies in their input history and the top-1 movie predicted by the Stage 1 LSTM model. These are combined with a standardized instruction template:

\textit{\textbf{Example prompt template:}}

\textit{Given the user's watched movies and the LSTM recommendation, recommend 3 more movies the user is likely to enjoy.}

\textit{- Watched: The Matrix, Inception, Fight Club, The Prestige, Memento}

\textit{- LSTM Suggests: Interstellar}

The target output is the next three movies the user actually watched, drawn from the held-out ground truth window:

\textit{\textbf{Example target output:}}

\textit{- The Lord of the Rings: The Fellowship of the Ring}

\textit{- Minority Report}

\textit{- The Bourne Identity}

The training uses the Hugging Face Transformers and peft library, where tokenization is performed using the respective tokenizer of the DeepSeek and Mistral tokenizers with a max sequence length of 512 tokens. LoRA is then applied to the attention projection layers with configuration of $r = 8$, $\alpha = 16$ and dropout of 0.1. Training is conducted for 3 epochs with a batch size of 2, gradient accumulation and mixed precision (FP16). The training loss is computed using the causal language modelling (CLM) objective, with the model learning to generate the output text token-by-token as a continuation of the prompt. After the fine-tuning, we then instructed the model to do free-form text generation tasks with the same prompt structure as discussed in Stage 2. This setup ensures that the LLM learns to map structured, context-rich prompts to coherent recommendations.

To further enhance the alignment between the LLM-generated recommendations and the user's inferred intent, we also implemented a re-ranking step based on semantic similarity. This step is designed to ensure that the movie titles proposed by the LLM remain consistent with the semantic trajectory captured by the LSTM in Stage 1. And also help prevent occasional LLM drift where generative models might return grammatically plausible but semantically irrelevant titles.

The process is as follows: each of the LLM generated movie titles (e.g., three per prompt) are encoded into a dense semantic vector using Sentence-BERT (SBERT, Reimers \& Gurevych, 2019). We use this pre-trained SBERT model (e.g., all-MiniLM-L6-v2) from the sentence-transformers library to generate 384-dimensional embeddings for each title. At the same time, we also convert the LSTM's Top-1 prediction into its corresponding title and generate SBERT embeddings for those as well. For each LLM-generated title, we compute the cosine similarity between its embedding and that of the LSTM's output. The three LLM-generated titles are then re-ranked in descending order by their similarity scores.

\textit{\textbf{Example re-ranking output:}}

\textit{Reranked by Sentence-BERT similarity:}

\textit{- Indiana Jones (1984) (sim=.8694)}

\textit{- Back to the Future (1985) (sim=0.5546)}

\textit{- The Goonies (1985) (sim=0.4752)}

Overall, the key innovation in DUALRec recommendation lies in its construction of an hybrid model, where the LSTM provides temporally-aware next-item prediction while the LLM generates semantically coherent recommendations guided by LSTM insights. Rather than explicitly modeling recommendation probabilities, the system leverages the LLM's learned text generation capabilities to produce contextually and semantically relevant movie suggestions.

\section{Experiments}

\subsection{Baselines}

To evaluate the performance of the proposed DUALRec framework, we adopt a benchmark strategy similar to that used in the Xu et al. (2025) research, as both models share a fundamental architecture, such as the use of LSTM for dynamic intent modelling and LLMs for personalized recommendation generation. But our implementation diverges in several key aspects, including prompt format (hard prompt instead of numerical soft prompts), model scale and the choice to do LoRA fine-tuning in our current study. However, despite these differences, the shared objective of making intent-adaptive next-item prediction does provide a basis for comparative benchmarking.

Given this architectural alignment, we used the baselines chosen by Xu et al. (2025) that allow for a more controlled performance comparison. The baselines span three major families of recommendation methods including traditional collaborative filtering and popularity-based models, deep learning-based sequential model and LLM or transformer-base models listed below:

• \textbf{Mostpop}: Recommend the most popular items based on users interaction history.

• \textbf{SKNN} (Jannach \& Ludewig, 2017): A memory-based recommendation model that identifies session-level item similarities using nearest-neighbor search, recommending items based on their co-occurrence with previously interacted items in the same session.

• \textbf{NARM} (Li et al., 2017): An RNN-based recommendation model enhanced with an attention mechanism, which learns to prioritize key items from a user's session history, effectively modeling both long-term and short-term user intent.

• \textbf{FPMC} (Rendle et al., 2010): Combines matrix factorization techniques with first-order Markov chain principles to effectively model sequential patterns in user-item interaction histories.

• \textbf{STAMP} (Liu et al., 2018): Emphasizes the significance of the most recent interaction through a novel short-term attention priority mechanism that balances current and historical preferences.

• \textbf{GCE-GNN} (Wang et al., 2020): Implements a dual-graph architecture that constructs both localized session graphs and global item-transition patterns to enrich item representations and session intent modeling.

• \textbf{MCPRN} (Wang et al., 2019): Employs a multi-channel purpose-routing network that identifies and processes multiple concurrent user intentions within a single session.

• \textbf{HIDE} (Li et al., 2022): Introduces a hierarchical intent disentanglement framework that decomposes item embeddings into distinct purpose-specific components to capture diverse user motivations.

• \textbf{Atten-Mixer} (Zhang et al., 2023): Deploys a multi-granular attention mixing architecture that identifies and processes continuous user intentions at various temporal resolutions.

• \textbf{UniSRec} (Hou et al., 2022): Presents a transfer learning framework that utilizes textual item descriptions to generate cross-domain representations, enabling recommendations across different item catalogs.

• \textbf{NIR} (Wang \& Lim, 2023): A language model-based approach that formulates next-item recommendation as a zero-shot prompting task, leveraging pre-trained LLMs to generate predictions without task-specific fine-tuning.

\subsection{Evaluation Matrix}

To assess the performance of the recommendation model, we employ standard metrics that are often used to assess recommender models (Xu et al., 2025; Wang \& Lim, 2023), including Hit Rate at n (HR@(1,5)) and Normalized Discounted Cumulative Gain at n (NDCG@(1,5)). HR@n measures the proportion of test cases where the relevant item appears within the top-n recommendations, calculated as the number of hits in top-k divided by the total number of test cases. NDCG@n evaluates the ranking quality by giving higher weight to correct predictions that appear earlier in the recommendation list.

In addition to ranking accuracy metrics, we also use Genre Jaccard Similarity to assess the semantic understanding of the model's recommendations. This metric compares the overlap of genre tags between the recommended items and the actual items the user watched next. By doing so, we evaluate how well each DUALRec variant preserves genre-level consistency, providing a complementary perspective on whether the generated movie recommendations align with the thematic preferences reflected in the user's true viewing behavior. Specifically, we compute the Jaccard similarity between the top-1 recommended movie and the user's actual next movie, focusing on shared genre labels to quantify semantic relevance.

\section{Result and Analysis}

This part represents the comprehensive experimental result and analysis of the current model, evaluating both the individual components and the integrated hybrid architecture. It will include the evaluation of LSTM model training performance, LLM fine-tuning effectiveness and comparative analysis against baseline models.

\subsection{LSTM Model Training Result}

The multimodal LSTM was trained for 10 epochs using the ML-1M dataset with the preprocessing pipeline described in Section 4.2. As shown in Figure 3, LSTM demonstrated a learning progression and convergence behaviour.

\begin{figure}[htbp]
\centering
\includegraphics[width=\linewidth]{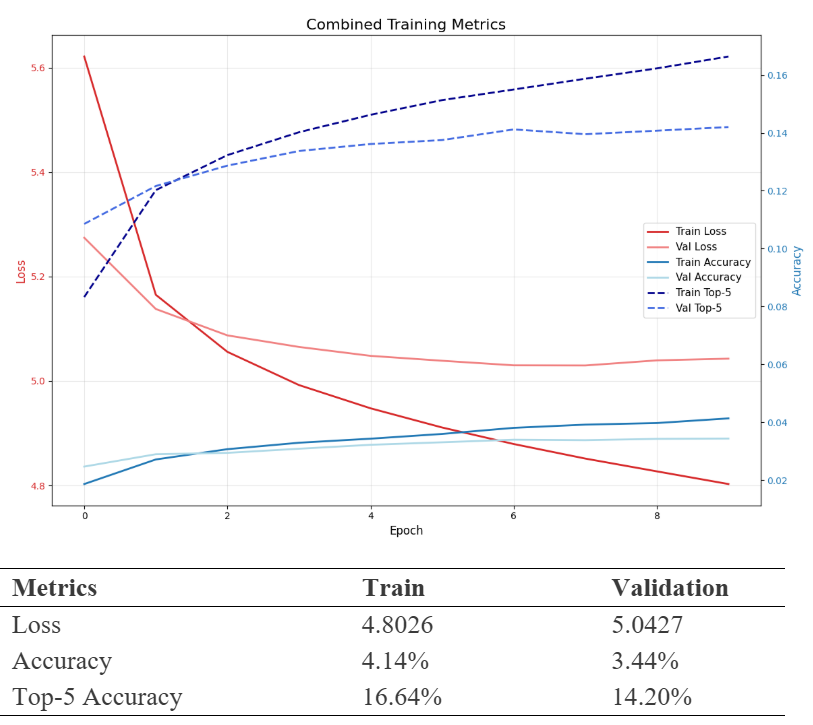}
\caption{Performance of LSTM model training with ML-1M dataset.}
\label{fig:lstm_training}
\end{figure}

For loss convergence, both training and validation loss exhibit clear convergence patterns with training loss decreasing from initial values of 5.63 to 4.80 by epoch 10. The validation loss follows a similar trend to stabilize around 5.04, which indicated an effective learning without overfitting. So the convergence pattern indicated that the model has successfully captured the underlying temporal patterns of user movie interactions.

For accuracy progression, the model demonstrates a steady improvement in prediction accuracy through training. The training accuracy increases from approximately 2.5\% to 4.14\% while the validation accuracy reaches 3.44\%. Although those accuracy values may appear modest, they are performed under the challenging nature of next-item prediction, where the model must select from 1000 candidate movies.

For the top-5 accuracy performance, it measures the percentage of cases where the correct next movie appears among the model's top-5 predictions, representing whether the model successfully identifies the actual user choice. The model shows a substantial improvement with reaching 16.64\% on training data and 14.20\% on validation set. This metric is particularly relevant for recommendation systems as users typically consider multiple suggested items rather than requiring the perfect top-1 predictions.

\subsection{DUALRec Variant Performance Analysis and LLM Fine-tuning}

As shown in Figure 4 and Figure 5(a), the DUALRec Mistral 7B fine tuned variant achieves the strongest performance among all four DUALRec configurations, and all 11 baselines with HR@1 and NDCG@ 1 of 0.1847. And its HR@5 performance of 0.2214, and NDCG@5 of 0.2078 although lower than top-performing baselines, it still demonstrated the model's ability to provide relevant recommendations within the top-5 suggestions. Compared to the top-5 accuracy of standalone LSTM models (0.1664 and 0.1420 respectively), the addition of an LLM improves recommendation accuracy, highlighting the benefit of combining sequential modeling with semantic reasoning. 

\begin{figure}[htbp]
\centering
\includegraphics[width=\linewidth]{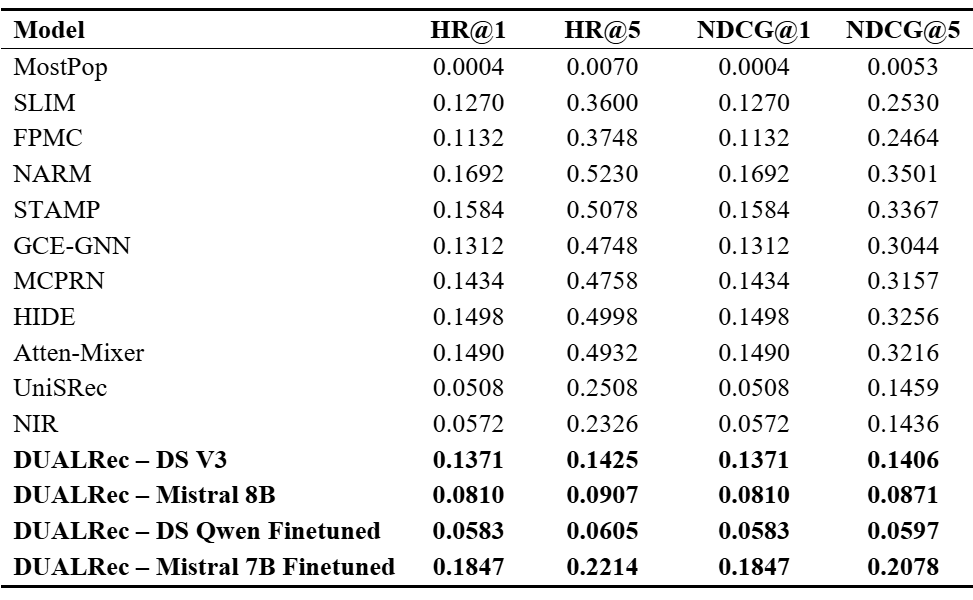}
\caption{Performance comparison of DUALRec variants versus baselines.}
\label{fig:performance_comparison}
\end{figure}

The DUALRec DeepSeek V3 variant showed moderate performance, with HR@1 and NDCG@1 scores of 0.1371, which is higher than 6 out of the 11 baselines. In contrast, the DUALRec Mistral 8B variant performed relatively poorly, with a score of only 0.0810 on the same metrics. The DUALRec DeepSeek Qwen variant exhibits the lowest performance among the four DUALRec variants, with HR@1 of 0.0583 and HR@5 of 0.0605. These results are only comparable to the NIR and UniSRec baselines, suggesting that choosing DeepSeek Qwen as an DeepSeek V3 alternative choice might not be as well suited for the current recommendation task.

\begin{figure}[htbp]
\centering
\includegraphics[width=\linewidth]{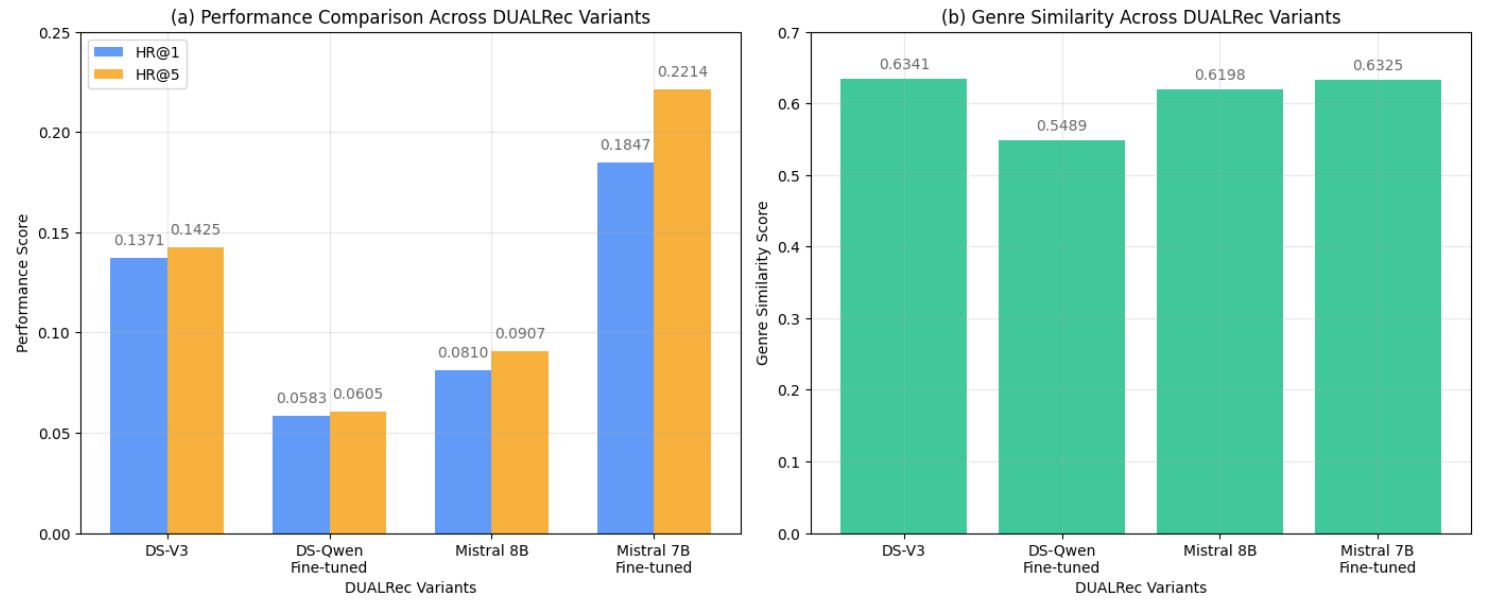}
\caption{(a) is the performance comparison across the choice of LLMs. (b) is the genre similarity comparison of the LLMs predicted next movies.}
\label{fig:genre_similarity}
\end{figure}

As shown in Figure 5(b), the DeepSeek V3 model achieves the highest genre similarity score of 0.6341, indicating strong semantic alignment with user preference even without fine-tuning. The Mistral 7B fine tuned variant closely follows a score of 0.6325, further confirming its ability to generate both contextual and thematic relevant movie recommendations. In contrast, the DeepSeek-R1-Distill-Qwen-1.5B variant yields a lower score of 0.5489, reflecting its relatively weaker performance in capturing general level  intent. These results confirmed that the DUALRec model architecture, especially with powerful LLM backbones like DeepSeek V3 and Mistral 7B fine tuned models, can not only predict correct next movie titles, but also could generate recommendations that are thematically aligned with actual user interests. 

\section{Discussion}

This study proposed DUALRec, a hybrid recommendation architecture that integrated a multi-modal LSTM-based sequential predictor with a prompt-based Large Language Model (LLM) for natural language reasoning. By combining temporal user behaviour with contextual inference, DUALRec aims to produce personalized and semantically coherent movie recommendations.

Among the DUALRec variants evaluated, a clear performance gap was observed between DeepSeek V3, DeepSeek-R1-Distill-Qwen-1.5B (fine-tuned) and Mistral 7B (fine-tuned). Notably, DeepSeek V3 despite not being fine-tuned, still outperformed DeepSeek Qwen finetuned in both accuracy and genre similarity. Conversely the relative poor performance of the fine-tuned DeepSeek Qwen variant may reflect overfitting or limited alignment between the pretraining corpus and recommendation domain. Also, while the model parameters between DeepSeek V3 and Mistral 7B diverge significantly (685B vs. 7B), the mistral 7B fine tuned model still achieved superior performance over the DeepSeek V3 model. Thus we could see that in recommendation tasks, especially those that are based on structured prompts, natural language processing ability alone might not guarantee strong performance, but the semantic alignment with both the input domain and output style might lead to successful recommendations.

The Mistral 7B fine tuned model achieved the highest HR@1 and NDCG@1 among all DUALRec variants and baselines, while also maintaining strong genre alignment; this could be explained by the Mistral's architecture. As it is designed for instruction following, dense casual attention and sliding window attention (Mistral AI, 2024a), which collectively enhance its ability to understand and extend prompts with high fidelity. Also with its smaller size of 7b parameters, as compared to DeepSeek V3 of 685b parameters, also facilitates faster adaptation with LoRA and higher computational efficiency. Besides, the large performance gap between the Mistral 8B and Mistral 7B LoRA fine-tuned variants have shown that although Mistral 8B is technically more recent and slightly larger model (Mistral AI, 2024b), its underperformance compared to the fine-tuned Mistral 7B highlights the importance of task-specific adaptation and fine tuning feasibility.

Another notable observation in our experiment is that the HR@5 and NDCG@5 scores did not significantly increase compared to HR@1 and NDCG@1 across all DUALRec variants. This pattern deviates from conventional baselines like NARM (Li et al., 2017) or STAMP (Liu et al., 2018), where top-K metrics typically show much higher recall as K increases. And this could be attributed to the semantic generation behaviour of LLMs. As language models are trained with maximum likelihood estimation (MLE) objective, which encourages them to generate high-probability and semantically cohesive outputs (Holtzman et al., 2020), not ranking diversity. Thus, leads to low diversity across the top-k outputs, which constrains performance improvements beyond the first-ranked item.

In contrast, conventional deep recommendation models, especially ranking-based architectures included in our baselines, are optimized to maximize discriminative spread across a candidate set, allowing them to retrieve a border range of items as K increases (Rendle et al., 2009; He et al., 2017; Kang \& McAuley, 2018). Together, we could see that while LLMs in DUALRec generate semantically relevant recommendations, they may still require explicit mechanisms to enhance recommendation diversity.

\section{Limitations and Conclusion}

Overall, DUALRec demonstrates the effectiveness of combining sequential modeling with LLM-based reasoning in a hybrid recommendation framework. However, the current study also has limitations. First, the model is trained and evaluated solely on the MovieLens 1M dataset, which represents a narrow movie domain, and user preferences in other domains (e.g., music, shopping, education) may differ significantly in data structure and variability. Future research should extend the DUALRec framework to diverse and multimodal datasets, incorporating features such as images, video clips, musics, or user-generated reviews, which can enrich the prompt and allow for more immersive personalization of the generated recommendations. Also, incorporating multimodal LLMs (e.g., DeepSeek Janus, GPT-4) with videos or picture processing abilities could enable more variety in recommendations.

Additionally, the current model is trained on offline, static data, which limits its responsiveness in real-world environments. In practice, recommender systems must scale to millions of users and adapt in real time to shifting preferences. Future research could focus on optimizing the DUALRec architecture for large-scale, real-time inference, while incorporating incremental or online learning algorithms to update user representations and LLM prompts dynamically as new interaction data becomes available.

In summary, this study demonstrates the feasibility and effectiveness of combining temporal sequence modeling with large-scale language understanding for recommendation. Future iterations of DUALRec can benefit from broader datasets, multimodal features, and real-time adaptation techniques to become even more responsive, personalized, and production-ready.

% Include the separate references file
% references.tex - Separate reference file
% This file contains manually formatted references in simple text format

\section*{References}

\noindent
Bakhshizadeh, M. (2024). Supporting knowledge workers through personal information assistance with context aware recommender systems. In \emph{Proceedings of the 18th ACM Conference on Recommender Systems} (pp. 1296-1301).

\noindent
Cheng, H.-T., Koc, L., Harmsen, J., Shaked, T., Chandra, T., Aradhye, H., Anderson, G., Corrado, G., Chai, W., Ispir, M., Anil, R., Haque, Z., Hong, L., Jain, V., Liu, X., \& Shah, H. (2016). Wide \& deep learning for recommender systems. In \emph{Proceedings of the 1st Workshop on Deep Learning for Recommender Systems} (pp. 7-10).

\noindent
Covington, P., Adams, J., \& Sargin, E. (2016). Deep neural networks for YouTube recommendations. In \emph{Proceedings of the 10th ACM Conference on Recommender Systems} (pp. 191-198).

\noindent
DeepSeek-AI. (2024a). DeepSeek: Scaling large language models with compute-optimal training and serving. \emph{arXiv preprint arXiv:2401.02954}.

\noindent
DeepSeek-AI. (2024b). DeepSeek-V3 [Pretrained language model]. Hugging Face. \url{https://huggingface.co/deepseek-ai/DeepSeek-V3}

\noindent
DeepSeek-AI. (2024c). DeepSeek-R1-Distill-Qwen-1.5B [Pretrained language model]. Hugging Face. \url{https://huggingface.co/deepseek-ai/DeepSeek-R1-Distill-Qwen-1.5B}

\noindent
Donkers, T., Loepp, B., \& Ziegler, J. (2017). Sequential user-based recurrent neural network recommendations. In \emph{Proceedings of the Eleventh ACM Conference on Recommender Systems} (pp. 152–160).

\noindent
Gao, C., Lei, W., He, X., de Rijke, M., \& Chua, T.-S. (2021). Advances and challenges in conversational recommender systems: A survey. \emph{AI Open}, 2, 100-126.

\noindent
Harper, F. M., \& Konstan, J. A. (2015). The MovieLens datasets: History and context. \emph{ACM Transactions on Interactive Intelligent Systems}, 5(4), 1–19.

\noindent
He, X., Liao, L., Zhang, H., Nie, L., Hu, X., \& Chua, T.-S. (2017). Neural collaborative filtering. In \emph{Proceedings of the 26th International Conference on World Wide Web} (pp. 173-182).

\noindent
Hidasi, B., Karatzoglou, A., Baltrunas, L., \& Tikk, D. (2015). Session-based recommendations with recurrent neural networks. \emph{arXiv preprint arXiv:1511.06939}.

\noindent
Holtzman, A., Buys, J., Du, L., Forbes, M., \& Choi, Y. (2020). The curious case of neural text degeneration. In \emph{Advances in Neural Information Processing Systems 33}.

\noindent
Hou, Y., Mu, S., Zhao, W. X., Li, Y., Ding, B., \& Wen, J.-R. (2022). Towards universal sequence representation learning for recommender systems. In \emph{Proceedings of the 28th ACM SIGKDD Conference on Knowledge Discovery and Data Mining} (pp. 585-593).

\noindent
Hu, E., Li, Y., Shen, Y., Wang, S., Wallis, P., Wang, L., Allen-Zhu, Z., \& Chen, W. (2021). LoRA: Low-rank adaptation of large language models. \emph{arXiv preprint arXiv:2106.09685}.

\noindent
Hugging Face. (2023). Parameter-efficient fine-tuning (PEFT). Hugging Face Blog. \url{https://huggingface.co/blog/peft}

\noindent
IBM. (2024). Recommendation engine. Retrieved July 16, 2025, from \url{https://www.ibm.com/think/topics/recommendation-engine}

\noindent
Jagerman, R., Markov, I., \& de Rijke, M. (2019). When people change their mind: Off-policy evaluation in non-stationary recommendation environments. In \emph{Proceedings of the Fifth ACM International Conference on Web Search and Data Mining} (pp. 447-455).

\noindent
Jannach, D., \& Ludewig, M. (2017). When recurrent neural networks meet the neighborhood for session-based recommendation. In \emph{Proceedings of the Eleventh ACM Conference on Recommender Systems} (pp. 306-310).

\noindent
Kang, W.-C., \& McAuley, J. (2018). Self-attentive sequential recommendation. In \emph{2018 IEEE International Conference on Data Mining} (pp. 197–206). IEEE.

\noindent
Li, J., Ren, P., Chen, Z., Ren, Z., Lian, T., \& Ma, J. (2017). Neural attentive session-based recommendation. In \emph{Proceedings of the 2017 ACM on Conference on Information and Knowledge Management} (pp. 1419-1428).

\noindent
Li, Y., Gao, C., Luo, H., Jin, D., \& Li, Y. (2022). Enhancing hypergraph neural networks with intent disentanglement for session-based recommendation. In \emph{Proceedings of the 45th International ACM SIGIR Conference on Research and Development in Information Retrieval} (pp. 1997-2002).

\noindent
Liu, Q., Zeng, Y., Mokhosi, R., \& Zhang, H. (2018). STAMP: Short-term attention/memory priority model for session-based recommendation. In \emph{Proceedings of the 24th ACM SIGKDD International Conference on Knowledge Discovery \& Data Mining} (pp. 1831-1839).

\noindent
Liu, W., Cheng, S., Zeng, D., \& Qu, H. (2023). Enhancing document-level event argument extraction with contextual clues and role relevance. \emph{arXiv preprint arXiv:2310.05991}.

\noindent
Liu, W., Zhou, L., Zeng, D., Xiao, Y., Cheng, S., Zhang, C., Lee, G., Zhang, M., \& Chen, W. (2024). Beyond single-event extraction: Towards efficient document-level multi-event argument extraction. \emph{arXiv preprint arXiv:2405.01884}.

\noindent
Mangrulkar, S., Gugger, S., Debut, L., Belkada, Y., Paul, S., \& Bossan, B. (2022). PEFT: State-of-the-art parameter-efficient fine-tuning methods [Computer software]. Hugging Face. \url{https://github.com/huggingface/peft}

\noindent
Mistral AI. (2024a). Mistral-7B-Instruct-v0.3 [Pretrained language model]. Hugging Face. \url{https://huggingface.co/mistralai/Mistral-7B-Instruct-v0.3}

\noindent
Mistral AI. (2024b). Mistral-8B-Instruct-2410 [Pretrained language model]. Hugging Face. \url{https://huggingface.co/mistralai/Ministral-8B-Instruct-2410}

\noindent
Musto, C., de Gemmis, M., Lops, P., Narducci, F., \& Semeraro, G. (2022). Semantics and content-based recommendations. In F. Ricci, L. Rokach, \& B. Shapira (Eds.), \emph{Recommender systems handbook} (pp. 251-295). Springer.

\noindent
Quadrana, M., Karatzoglou, A., Hidasi, B., \& Cremonesi, P. (2017). Personalizing session-based recommendations with hierarchical recurrent neural networks. In \emph{Proceedings of the Eleventh ACM Conference on Recommender Systems} (pp. 130–137).

\noindent
Reimers, N., \& Gurevych, I. (2019). Sentence‑BERT: Sentence embeddings using siamese BERT‑networks. In \emph{Proceedings of the 2019 Conference on Empirical Methods in Natural Language Processing and the 9th International Joint Conference on Natural Language Processing} (pp. 3982–3992). Association for Computational Linguistics.

\noindent
Ren, X., Wei, W., Xia, L., Su, L., Cheng, S., Wang, J., Yin, D., \& Huang, C. (2024). Representation learning with large language models for recommendation. In \emph{Proceedings of the ACM Web Conference 2024} (pp. 3464–3475).

\noindent
Rendle, S., Freudenthaler, C., Gantner, Z., \& Schmidt-Thieme, L. (2009). BPR: Bayesian personalized ranking from implicit feedback. In \emph{Proceedings of the Twenty-Fifth Conference on Uncertainty in Artificial Intelligence}.

\noindent
Rendle, S., Freudenthaler, C., \& Schmidt-Thieme, L. (2010). Factorizing personalized Markov chains for next-basket recommendation. In \emph{Proceedings of the 19th International Conference on the World Wide Web} (pp. 811–820).

\noindent
Sarwar, B., Karypis, G., Konstan, J., \& Riedl, J. (2001). Item-based collaborative filtering recommendation algorithms. In \emph{Proceedings of the 10th International Conference on the World Wide Web} (pp. 285–295).

\noindent
Sun, F., Liu, Q., Wu, S., Pei, Y., Lin, Z., \& Wang, L. (2019). BERT4Rec: Sequential recommendation with bidirectional encoder representations from transformer. In \emph{Proceedings of the 28th ACM International Conference on Information and Knowledge Management} (pp. 1441–1450). ACM.

\noindent
Wang, Q., Li, J., Wang, S., Xing, Q., Niu, R., Kong, H., Li, R., Long, G., Chang, Y., \& Zhang, C. (2018). Towards next-generation LLM-based recommender systems: A survey and beyond. \emph{ACM Computing Surveys}.

\noindent
Wang, S., Hu, L., Wang, Y., Sheng, Q. Z., Orgun, M., \& Cao, L. (2019). Modeling multi-purpose sessions for next-item recommendations via mixture-channel purpose routing networks. In \emph{International Joint Conference on Artificial Intelligence}. International Joint Conferences on Artificial Intelligence.

\noindent
Wang, Z., Wei, W., Cong, G., Li, X. L., Mao, X. L., \& Qiu, M. (2020). Global context enhanced graph neural networks for session-based recommendation. In \emph{Proceedings of the 43rd International ACM SIGIR Conference on Research and Development in Information Retrieval} (pp. 169-178).

\noindent
Wang, L., \& Lim, E.-P. (2023). Zero-shot next-item recommendation using large pretrained language models. \emph{arXiv preprint arXiv:2305.00962}.

\noindent
Wang, S., Zhang, X., Wang, Y., \& Ricci, F. (2024). Trustworthy recommender systems. \emph{ACM Transactions on Intelligent Systems and Technology}, 15(4), 1–20.

\noindent
Xu, X., Zhang, Y., Wang, L., Chen, J., \& Li, H. (2025). Enhancing user intent for recommendation systems via large language models. \emph{Preprints}.

\noindent
Yu, P., Cui, V. Y., \& Guan, J. (2021). Text classification by using natural language processing. \emph{Journal of Physics: Conference Series}, 1802(4), 042010. IOP Publishing.

\noindent
Yu, P., Xu, X., \& Wang, J. (2024). Applications of large language models in multimodal learning. \emph{Journal of Computer Technology and Applied Mathematics}, 1(4), 108-116.

\noindent
Zhang, P., Guo, J., Li, C., Xie, Y., Kim, J. B., Zhang, Y., Xie, X., Wang, H., \& Kim, S. (2023). Efficiently leveraging multi-level user intent for session-based recommendation via atten-mixer network. In \emph{Proceedings of the Sixteenth ACM International Conference on Web Search and Data Mining} (pp. 168–176). Association for Computing Machinery.
\end{document}